\documentclass[fleqn,10pt,twocolumn]{wlscirep}
\usepackage[utf8]{inputenc}
\usepackage[T1]{fontenc}

\usepackage{amsbsy}
\usepackage{bm}
\usepackage{fixmath}
\usepackage{xfakebold}

\usepackage[table,xcdraw]{xcolor}
\usepackage{colortbl}
\usepackage{graphicx}
\usepackage{dcolumn}
\usepackage{bm}
\usepackage{multirow}
\usepackage{array}
\usepackage{subfigure}
\usepackage[export]{adjustbox}
\usepackage{tikz}
\usetikzlibrary{calc}
\newcolumntype{P}[1]{>{\centering\arraybackslash}p{#1}}
\usepackage{siunitx} 
\usepackage{booktabs}
\usepackage{adjustbox}
\usepackage[english]{babel}
\usepackage{listings}
\usepackage[utf8]{inputenc}
\usepackage[T1]{fontenc}
\usepackage{float}
\usepackage[utf8]{inputenc}
\usepackage{amsmath}
\usepackage{algorithm}
\usepackage{algpseudocode}
\usepackage{amsfonts}
\usepackage{xcolor}
\usepackage{caption}
\usepackage{tabularx}
\usepackage{setspace}
\usepackage{mathrsfs}

\title{Adaptive workforce exploration in complex productivity landscapes}

\author[1,2]{Mateus F. B. Granha}
\author[3,1,2]{Igor V. G. de Oliveira}
\author[1,2]{Andr\'e L. M. Vilela}
\author[4]{Chao Wang}
\author[2]{Paulo R. A. Campos}
\affil[1]{F\'isica de Materiais, Universidade de Pernambuco, Recife, PE 50720-001, Brazil}
\affil[2]{Departamento de F\'isica, Universidade Federal de Pernambuco, Recife, PE 50670-901, Brazil}
\affil[3]{Instituto de F\'isica, Universidade de S\~ao Paulo, S\~ao Paulo, SP 05314-970, Brazil}
\affil[4,*]{College of Economics and Management, Beijing University of Technology, Beijing, 100124, China}

\affil[*]{Electronic address: chaowanghn@vip.163.com}

\begin{abstract}
Specialization and task allocation enhance efficiency and innovation across diverse systems, from biological organisms to socioeconomic institutions. The evolution of task distribution and its influence on organizational productivity encapsulate the dynamics between task dependencies and adaptive strategies. We explore the organizational division of labor, inspired by the NK model of rugged landscapes, which is widely applied in evolutionary biology, and incorporate interdependencies among the attributes of technical experts within an organization. Our model considers two types of employees characterized by their task allocation strategies: specialists, who are permanently assigned to a single task, and generalists, who stochastically select a task at each time step. We investigate how the ruggedness of the productivity landscape, shaped by task interdependency, affects the organization's capacity to optimize labor division and meet market demands. Using group selection algorithms, we reveal the emergence of nonlinear adaptive dynamics, providing insights into how companies can adapt their strategies to meet market demands and foster innovation.
\end{abstract}

\begin{document}
\flushbottom
\maketitle
\thispagestyle{empty}

\section{Introduction}

Strategic division of labor is critical for driving efficiency and fostering innovation in the modern business environment \cite{UlrichNature2018, YangTechFSC2024}. Organizations can respond competitively to market and societal demands by optimizing how they divide and coordinate tasks, thereby maximizing socioeconomic outcomes \cite{ChakrabartiRDManag1989}. However, the complex dependencies among tasks pose significant challenges to achieving optimal productivity \cite{SaavedraJAP1993}. Identifying effective task allocation strategies requires an understanding of the intricate relationships between individual performance and the emergent dynamics emerging from interactions among agents and tasks. The complexity of these interactions underscores the need for quantitative frameworks that account for the interplay among specialization, coordination, and adaptive behavior in organizational settings.

Studies regarding division of labor, economics, and management benefit from investigations that use the concept of fitness landscapes \cite{CamposPhysicaA2019, CamposEcolComp2024}. The concept of the fitness landscape, introduced by Sewall Wright \cite{Wright1932}, comprises a powerful tool for representing and visualizing adaptive processes. Accordingly, evolutionary adaptation consists of an uphill climb on a fitness landscape, seeking the highest fitness peaks. The fitness landscape is a genotype-fitness mapping that assigns a fitness value to each genotype configuration. At the individual level, the genotype may correspond to a set of employees' attributes for performing specific tasks, whereas fitness serves as a proxy for their performance. In this framework, organizational adaptation can be interpreted as a search for high-performance configurations on a landscape whose structure determines the difficulty of finding globally optimal solutions.

\begin{figure*}[ht]
    \centering
    \includegraphics[width=0.70\linewidth]{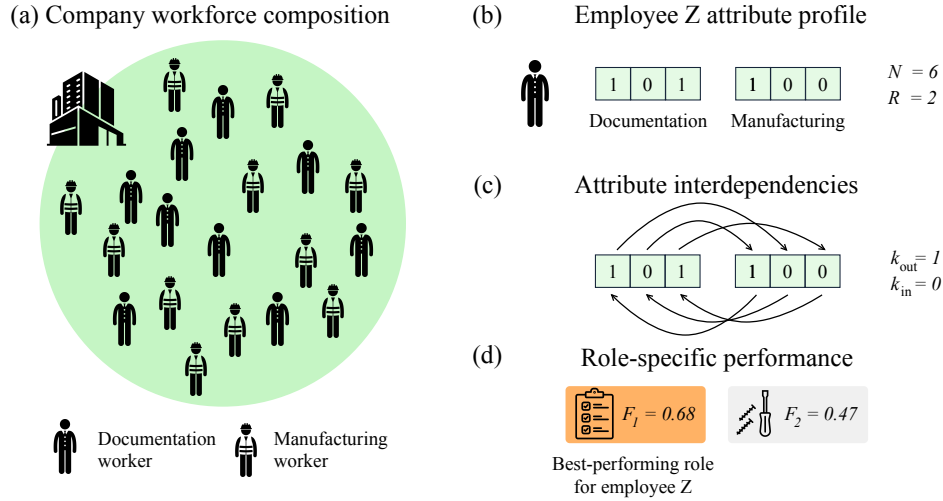}
    %\vspace{-1cm}
    \caption{Illustration of a company comprised of two departments: documentation and manufacturing. (a) Company workforce composition comprising $20$ employees. (b) Visual representation of the trait structure of employee Z, where each task associated with a department has three attributes that the employee may or may not have, totaling $N = 6$ attributes per employee. (c) Arrows indicate interdependency between these attributes. In this example, there are no internal interdependencies $k_{\text{in}} = 0$ and each attribute depends on one attribute of another task, $k_{\text{out}} = 1$. (d) In this case, the first task has the highest performance (fitness). Hence, we highlight documentation as the best-performing task for employee Z.}
    \label{fig: traits illustration}
\end{figure*}

A distinguished class of theoretical fitness landscape models is the NK model \cite{kauffman1987towards}. A critical feature of the NK model is the ability to tune the degree of ruggedness of the fitness landscape, thereby creating a family of landscapes with different structural properties. Modern applications of the NK model comprise evolutionary biology \cite{CamposPhysicaA2002, FragataTREE2019}, optimization problems \cite{WrightIEEE2000, Weinberger1996}, innovation policies and ecosystems \cite{Desmarchelier2013, Luo2018, Bassart2026, GancoRP2017}. The model also has utility in economics for studying technological innovation and industry dynamics, revealing how corporations navigate demanding technological landscapes to achieve breakthroughs \cite{auerswald2000}. In management and organizational theory, the NK model is an assertive tool for examining strategic decision-making and organizational design, helping organizations balance exploration and exploitation, adapt to changing environments, and optimize performance \cite{Levinthal1997, Rivkin2003}. Additionally, the model has applications in understanding how the social network structure affects information dissemination and the evolution of cultural traits in social systems \cite{LazerAdminSci2007}.

The standard NK model depicts a system with $N$ components, each with a state variable that assumes one of two values at any given time, $a_{\ell} = 0$ or $1$, for $\ell = 1, \ldots, N$. The contribution of each component $i$ to the overall fitness depends on its state and the state of $k$ randomly chosen components. The parameter $k$ is the epistasis parameter, which tunes the ruggedness of the landscape. In a broader scenario, $k$ represents the level of complexity and interdependency within the system, which increases with higher values of $k$ \cite{Levinthal1997, MaEuroJOperRes2005, frenken2006technological, billinger2014search, celo2015mnc, GancoRP2017, ganco2020rugged, li2022computational}. The resulting rugged landscape has numerous local maxima, increasing the sophistication of achieving optimal performance \cite{weinberger1991local}.

In biological systems, the components of the NK model map genes inherently: the effect of a given gene (or allele) on the phenotype often depends on the state of many other genes, a phenomenon known as epistasis. Each gene occupies a specific locus (a physical position on a chromosome), and interactions among loci generate complex, path-dependent evolutionary trajectories on a rugged fitness landscape. A concrete example is protein folding, where the three-dimensional structure and function of a protein emerge from the joint interactions among many amino acids encoded by different genes; small substitutions can have strongly context-dependent effects, making the fitness landscape rough and difficult to navigate \cite{kauffman1993origins}. 

In corporate settings, components represent distinct attributes or traits required to perform various tasks, each affecting and being affected by several others. In Fig. \ref{fig: traits illustration}, we illustrate in panel (a) a company composition of $20$ employees, and in panel (b) an individual Z with six traits that can be applied in the operation of two roles, with respective performance indicators: documentation and manufacturing. In Fig. \ref{fig: traits illustration}(c), arrows represent dependencies between traits across different roles. For example, an employee with strong organizational skills in documentation may leverage this trait in multiple activities within the company, such as workflow planning and compliance management. In this sense, a single attribute can contribute simultaneously to the performance of other interconnected tasks, generating interdependencies across the organizational performance landscape. Fig. \ref{fig: traits illustration}(d) illustrates that the combination of employee Z attributes determines their performance score for each task, allowing them to execute the one for which they are best suited -- in this case, documentation.

Similar patterns of interdependence also characterize a broad range of socioeconomic systems, in which the success of each module depends on its interactions with others. For instance, marketing strategies depend on integrating product development, sales tactics, and effective customer feedback management. In supply chain management, the efficiency of a manufacturing process relies on the timely delivery of raw materials, coordination between production stages, and logistics for distributing final products. Each component must work in harmony to achieve optimal efficiency, and any disruption in one area can significantly impact the entire supply chain. In financial markets, the stability of an investment portfolio depends on the joint dynamics between different asset classes, market trends, and economic indicators. Investors must consider the correlations between these components to balance risk and return effectively.

\section{Patterns of Labor Allocation in Organizations}

\begin{figure*}[ht]
    \centering
    \includegraphics[width=1.0\linewidth]{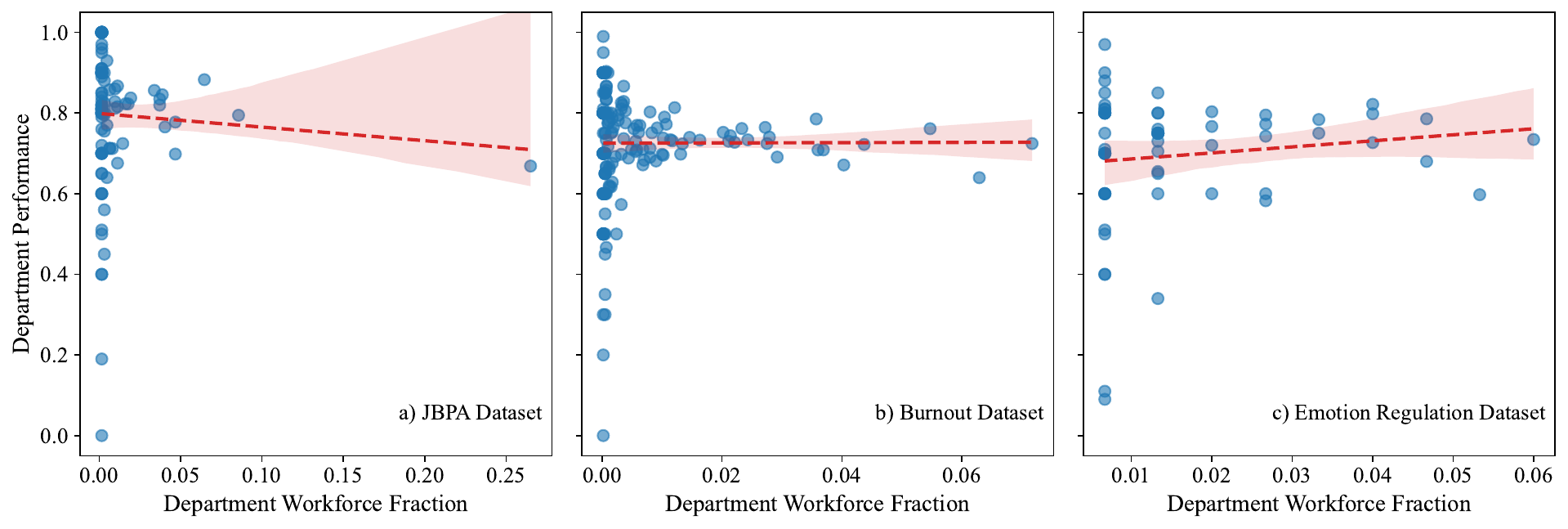}
    \caption{Relationship between department workforce fraction and normalized department performance across three empirical public sector datasets: a) JBPA Dataset, b) Burnout Dataset, and c) Emotion Regulation Dataset. Dashed red lines represent fitted linear regression models accompanied by $95\%$ confidence intervals (shaded regions). The linear regressions are nearly flat across all panels, indicating that department size is not significantly correlated with performance measures.}
    \label{fig:empirical_datasets}
\end{figure*}

Real organizations operate in heterogeneous task environments, unequal workforce distributions, and complex dependencies among activities. To illustrate this phenomenon, we examine three empirical datasets from public administration research repositories with distinct structures and operational profiles \cite{resh2020emotional, le2022asianamericans}. These datasets provide a real-world perspective on how labor is allocated across functions and how organizational performance varies across departments, highlighting patterns that inspire the simplified framework introduced later in this work.

The datasets record employee survey responses across municipal and local government departments. The first dataset, referred to as the JBPA (Journal of Behavioral Public Administration) Dataset \cite{resh2020emotional}, contains responses from $N = 645$ municipal workers distributed across $100$ departments. The underlying study examined public engagement, emotional labor, and recall bias in local government. The second and third datasets derive from a single research that examined burnout, emotion regulation, and workplace dynamics among local government staff \cite{le2022asianamericans}. The Burnout Dataset includes $N = 6071$ employees across $194$ departments and assessment of occupational stress, burnout dimensions, and organizational outcomes across public sector entities. The Emotion Regulation Dataset represents a specialized panel of $N = 1662$ workers across $62$ departments, collecting data on daily emotional regulation tactics, such as emotion suppression and amplification, as well as work-life boundaries during operational shifts.\footnote{Data provided by Harvard Dataverse.}

Across all datasets, performance was evaluated using employee ratings of overall department performance on an 11-point scale from $0$ to $10$, which we normalized to the range $0.0$ to $1.0$. We deliberately selected departmental performance ratings over individual self-evaluations because self-evaluations are prone to self-serving bias, leading to inflated ratings clustered near the maximum. Rating overall departmental output provides a more rigorous, critical, and realistic assessment, with greater statistical variance across units. Table~\ref{tab:dataset_summary} summarizes the sample size, number of departments, and average performance scores for all datasets.

\begin{table*}[htbp]
\centering
\caption{Summary of empirical datasets, including workforce size, total departments, and average department performance. We consider only departments that reported both workforce size and average performance.}
\label{tab:dataset_summary}
\begin{tabular}{lccc}
\toprule \midrule
Dataset & Employees & Departments & Average Performance \\
\midrule
JBPA & $645$ & $100$ & $0.7956$ \\
Burnout & $6071$ & $194$ & $0.7255$ \\
Emotion Regulation & $1662$ & $62$ & $0.6953$ \\
\bottomrule
\end{tabular}
\end{table*}

To examine how department size relates to performance, we calculate the workforce fraction for each department as the ratio of its employees to the total number of respondents in the dataset. Figure \ref{fig:empirical_datasets} shows the relationship between department workforce fraction and normalized performance across all datasets. The data reveals that the distribution of workers across departments is skewed. Most departments are small, accounting for less than $3\%$ of the total organizational workforce, exhibiting wide variation in performance. In contrast, larger operational departments, such as fire protection and public works, show narrower performance variation, converging near $0.70$ to $0.75$.

The linear regression lines (dashed red lines) with $95\%$ confidence intervals are nearly flat across all panels, indicating that department size does not have a simple linear relationship with perceived performance. Simply adding more personnel to a department does not increase its average performance score. As departments grow, coordination needs and task interdependencies increase, which can offset gains from higher workforce capacity. These observations suggest that organizational productivity depends on how tasks are structured and how workers' skills are allocated, rather than on workforce size alone.

These empirical patterns motivate a systematic analysis of how alternative labor division strategies influence organizational efficiency. Although real organizations feature many departments, we investigate the fundamental challenge of the division of labor by introducing a minimal two-role setting that illustrates a trade-off between specialization and generalization. We adopt a two-task framework to isolate the core mechanisms controlling the division of labor, specifically demand asymmetry, skill alignment, and the interdependencies that shape the underlying performance landscape. We reduce the problem to two tasks to systematically investigate how organizations can allocate specialists and generalists in response to changing conditions while maintaining computational tractability. This framework captures nontrivial phenomena, such as the non-monotonic dependence of labor division on the generalization threshold and the dual role of task interdependencies, which are expected to persist and expand in multi-task settings.

The concept of division of labor, in which tasks are allocated according to workers' skill and expertise, is fundamental to organizational theory and economics. Specialization allows employees to develop greater task proficiency, often leading to higher efficiency and productivity. At the same time, modern organizations operate in dynamic environments that require flexibility, rapid adaptation, and responsiveness to changing demands. Consequently, organizations must continuously balance the benefits of specialization against the adaptability of more general workers. Understanding how this balance emerges and influences organizational performance is therefore crucial to achieving sustained productivity and innovation \cite{march1991, adler1999}.

Our work extends the NK model by incorporating a group search algorithm to investigate how specialists and generalists emerge and coexist within modern socioeconomic institutions. Using a quantitative framework that combines adaptive learning, innovation, and interdependent performance landscapes, we analyze how organizations allocate workers across tasks and how these allocation strategies influence productivity. The results identify the conditions under which specialization, generalization, or mixed workforce structures maximize organizational performance, while highlighting the role of task interdependencies in shaping productivity landscapes.

\section{Modeling Design}

\subsection{NK model for organizational performance measures}
We represent the professional profile of the employee in an organization, comprising $N$ attributes $\vec{a} = \{a_1, a_2, \ldots, a_N\}$, each of whose contributions is influenced by $k$ other traits. The parameter $k$ quantifies the degree of interdependence among traits. Each trait may assume one of two possible states at any given time, $a_{\ell} = 0$ or $a_{\ell} = 1$, representing the absence or presence of the corresponding trait, respectively. The average performance in a given task, or role, associated with a number of traits $N$, for an employee with a professional profile $\vec{a}$ is
\begin{equation}
F(\vec{a}, k, N) = \frac{1}{N} \sum_{\ell \in \{ N \}} f_{\ell}(a_{\ell}, a_{\ell 1}, a_{\ell 2}, \ldots, a_{\ell k}),
\label{eq: fitness}
\end{equation}
where $a_{\ell}, a_{\ell 1}, a_{\ell 2}, \ldots, a_{\ell k}$ represent the states of the traits which are dependent on other traits through $k$ connections in the set $\{N\}$. The function $f_{\ell}(a_{\ell}, a_{\ell 1}, a_{\ell 2}, \ldots, a_{\ell k})$ is drawn from a uniform distribution in the interval $[0, 1]$, and represent the performance contribution of trait $\ell$ to the overall employee's performance. The performance associated with the role emerges from the combined effects of multiple interdependent employee attributes.

The pair of parameters $N$ and $k$ defines a rugged fitness landscape, and tuning their values critically influences its shape. In particular, when $k = 0$, the landscape is smooth and has a single maximum, and typical local search mechanisms will invariably converge to this global maximum. Conversely, increasing $k$ leads to a progressive increase in the number of local maxima, resulting in a rugged landscape and increasing the likelihood that the system becomes trapped in local optima \cite{weinberger1991local, kauffman1987towards}. 

\subsection{Division of labor: two role framework}

Our model presents a variant of the standard NK model to investigate the organizational structure of modern organizations. Thus, we depict employees as a sequence of $N$ attributes or abilities influenced by the interdependence parameter $k$. We obtain a performance landscape by assigning a performance score to each task, and evaluate each employee's performance by dividing their attributes into contributions for different organizational roles. In the two-task framework considered here, for a sequence of size $N$, we partition the attributes into two role domains: the first $N_1 = N/2$ attributes contribute to performance in role $1$, whereas the remaining $N_2 = N/2$ contribute to performance in role $2$. However, the contribution of a given attribute to a task is not determined solely by attributes within its own task domain; it can also be influenced by attributes associated with other tasks. Each attribute interacts with a total of $k$ other attributes. Of these, $k_{\rm in}$ belong to the same task domain, whereas $k_{\rm out}$ belong to the other domain, such that $k = k_{\rm in} + k_{\rm out}$.

The performance of an employee with a professional profile $\vec{a}$ in role $r$ is defined as the average contribution of the attributes associated with that task:
\begin{equation}
F_{r}(\vec{a}, k, N_r) = \frac{1}{N_r} \sum_{\ell\, \in \, \{N_r\}} f_{\ell}(a_{\ell}, a_{\ell 1}, a_{\ell 2}, \ldots, a_{\ell k}),
\label{eq: fitness by task}
\end{equation}
where $N_r$ is the number of attributes assigned to role $r$ and $\{N_r\}$ is the corresponding set of attributes.

The workforce consists of two types of employees: specialists and generalists. Specialists are assigned exclusively to the task at which they perform best, whereas generalists can perform either task and are dynamically allocated according to organizational demand. We classify an employee as a generalist whenever the difference between their performances in the two tasks satisfies
\begin{equation}
|F_1(\vec{a}, k, N_1) - F_2(\vec{a}, k, N_2)| < \theta.
\end{equation}
where $\theta$ is the generalization threshold. Employees whose performance difference exceeds this threshold are classified as specialists and assigned to their highest-performing task. Consequently, $\theta$ controls the degree of workforce flexibility, with $\theta = 0$ corresponding to the limiting case in which all employees are specialists.

To find the optimal organizational configuration, we explore how generalists are allocated in response to the organization's unmet demand. We define a labor-shortage measure that compares the market demand $\beta_r$ with the fraction of specialists assigned to role $r$, for a given organizational configuration $c$, as
\begin{equation}
    U_{rc} = \begin{cases}
        \beta_r - s_{rc},  & \text{if } s_{rc} < \beta_r, \\
        0, & \text{if } s_{rc} \geq \beta_r,
    \end{cases}
\end{equation}
where $s_{rc}$ denotes the fraction of specialists performing task $r$ in the candidate workforce design $c$. Thus, a labor shortage $U_{rc}$ is positive only when the current specialist workforce is insufficient to satisfy the corresponding demand, and it vanishes once that demand has been met. For a total number of organizational tasks $R$, generalist agents are assigned to tasks with probabilities proportional to their corresponding labor shortages,
\begin{equation}
    p_{rc} = \frac{U_{rc}}{\sum_{\, r \, =\, 1}^{\, R} U_{rc}} = \frac{U_{rc}}{U_{1c} + U_{2c}},
\end{equation}
since we consider $R = 2$. The total workforce allocation fraction to role $r$ in organizational allocation $c$ is
\begin{equation}
	n_{rc} = s_{rc} + g_{rc}.
\end{equation}
in which $g_{rc}$ denote the fraction of generalists assigned to role $r$ and organizational desing $c$. Generalists act as a flexible component that is preferentially directed toward tasks experiencing the largest workforce deficits.

Once all employees have been assigned to tasks, we evaluate the productivity of the organizational configuration $c$ based on its ability to satisfy market demand. To account for both workforce allocation and the trade-off between flexibility and specialization, we define the unnormalized productivity as
\begin{equation}
W_{c}(n_{1c} \,, \ldots, \, n_{Rc}) = \prod_{r \, =\, 1}^{R} n_{rc}^{\beta_r} (1 - \epsilon)^{g_{rc}} = n_{1c}^{\beta_1} \, n_{2c}^{\beta_2} \, (1 - \epsilon)^{g_{1c} + g_{2c}}.
\label{eq: productivity}
\end{equation}
The first term measures the degree to which workforce allocation matches market demand, $\beta_r$. The factor $(1-\epsilon)^{g_{1c} + g_{2c}}$ accounts for the reduced efficiency of generalists relative to specialists, where $\epsilon$ is the performance cost associated with generalization.

To standardize productivity analysis across diverse landscapes, we normalize the expression to its theoretical maximum. At this point, it is important to emphasize that maximum productivity is achieved when only specialists are present, $g_{rc} = 0$ and $n_{rc} = s_{rc}$, thereby allowing us to estimate its maximum value analytically. We maximize the organization's productivity under the constraints $\sum_{r \,=\, 1}^{R} s_{rc} = 1$ and $\sum_{r\, =\, 1}^{R}\beta_r = 1$ \cite{DuarteBehEcolSoc2012,CamposEcolComp2024}. The Lagrangian function for this optimization in a workforce configuration $c$ is constructed as follows
\begin{multline}
  L(s_{1c}, \ldots, s_{Rc}, \beta_1, \ldots, \beta_R, \lambda_R, \lambda_\beta) = \\ = \prod_{r\, =\, 1}^{R} s_{rc}^{\beta_r} - \lambda_R \left ( \sum_{r \, =\, 1}^{R} s_{rc} - 1 \right ) - \lambda_\beta \left ( \sum_{r \,=\, 1}^{R}\beta_r - 1 \right).
\end{multline}
Solving the system of equations $\vec{\nabla} L = \vec{0}$ implies that maximum productivity is achieved when the distribution of employees exactly matches the market demands, $s_{rc} = \beta_r \ \forall \ r$, yielding $W_{\mathrm{max}} = \prod_{i\, =\, 1}^{R} \beta_r^{\beta_r}$. 

For a general case, where $n_{rc} = s_{rc} + g_{rc}$, the normalized productivity of the $c$-th configuration $w_c$ is written as
\begin{multline}
w_{c}(n_{1c}, \ldots, n_{Rc}) = \\
= \frac{W_c(n_{1c} , \ldots, n_{Rc})}{W_{\mathrm{max}}} = \frac{\prod_{\, r\, =\, 1}^{\, R} n_{rc}^{\beta_r} (1-\epsilon)^{g_{rc}} }{\prod_{\, r\, =\, 1}^{\, R} \beta_r^{\beta_r}},
\label{eq: normalized productivity}
\end{multline}
where we represent the set of all configurations as $\mathbf{w} = (w_1, \ldots, w_C)$. Finally, we consider $C$ versions of a company's organizational design that evolve independently across the productivity landscape. Each version is characterized by the set of attributes present for employee $e$ under organizational design $c$, that is, $\vec{a}_{e, c} = \{a_{\ell}\}_{e, c}$. We measure the average organizational productivity of the system with $R$ roles at a given time step as
\begin{equation}
    P_{\, R} = \frac{1}{C} \sum_{c \, =\, 1}^{C} w_c(n_{1c} , \ldots, n_{Rc}).
\end{equation}

Following group selection, organizational designs with greater productivity tend to increase in occurrence. The distribution of employees across tasks should match consumer demand, and we quantify deviations from the optimal division of labor as
\begin{equation}\label{eq:nu}
\nu_r = \left | \frac{\beta_r - \langle{n}_{r}(t)\rangle_{c}}{\beta_r} \right |, \quad \textrm{with} \quad \nu = \textrm{max}(\nu_r).
\end{equation}
where $\langle{n}_{r}(t)\rangle_{c}$ denotes the average fraction of individuals performing role $r$ among all organization configurations $C$. We consider the largest role-division imperfection across all roles, $\nu = \textrm{max}(\nu_r)$.

\subsection{Workforce optimization}

We use a group-selection framework to investigate the conditions underlying the emergence of optimal division of labor. We associate groups that achieve better task partitioning with higher productivity. These configurations are more likely to be propagated to the next generation of workforce design, increasing the organization's overall performance in search of the optimal task distribution.

\begin{table}[h!]
\centering
\renewcommand{\arraystretch}{1.4}
\begin{tabular}{
    >{\hspace{3pt}}p{1.3cm}
    >{\hspace{3pt}}p{4.8cm} 
    >{\hspace{3pt}}p{1.2cm}
} 
\toprule \midrule
Variable & Description & Value \\ \noalign{\vspace{0.2cm}}
\midrule
$N$  & Number of attributes   & $16$ \\
$R$ & Number of roles & $2$ \\
$C$  & Number of organizational designs     & $100$ \\
$M$ & Organization size  & $50$ \\
$\mu$  & Attribute exploration probability   & $10^{-4}$ \\
$\eta$ & Number of independent samples & $10^4$ \\
$\tau$ & Relaxation time & $3 \times 10^4$ \\
$T$ & Simulation measuring time & $10^3$ \\
$t_{\text{skip}}$ & Sampling rate & $10$ \\
\bottomrule
\end{tabular}
\caption{Simulation parameters and numerical setup.}
\label{table: parameters}
\end{table}

Our model comprises $C$ versions of an organization's employee group, each consisting of $M$ individuals. We implement the standard Wright-Fisher process, in which organizational designs with higher productivity are more likely to be replicated in the next generation, while the total number of designs remains constant. Each newly generated organization initially copies the employee attribute configurations of the selected design, after which each employee may, with probability $\mu$, modify one randomly chosen attribute by flipping its binary state, $a_{\ell} = 0 \leftrightarrow 1$. This exploration mechanism introduces variation and allows the evolutionary search to explore alternative workforce configurations across the productivity landscape.

\section{Results}

\subsection{Simulation protocol}

Our model describes an organization comprising $M$ employees, each with $N$ attributes. Employees can perform one of $R$ organizational roles, with specialists following assignments aligned to their maximum performance and generalists tending to assume roles with the highest labor shortage. We initialize $C$ versions of the same organization with all employees having the same attribute configuration $\vec{a} = \{0\}$. We investigate the model's behavior across different combinations of the threshold parameter $\theta$ and the market demand parameter $\beta$.

At each generation, we first compute the normalized productivity of each organizational design $w_c$ using Eq. \eqref{eq: normalized productivity}. Organizational configurations evolve in line with productivity, such that highly productive organizations $c$ are more likely to persist and generate new organizational variants. The total number of configurations $C$ remains constant over time. 

New organizational variants are generated by preferentially replicating highly productive organizations. Specifically, organization $c$ is selected with probability proportional to its productivity $w_c$, after which its workforce composition is copied to form a new candidate organization. During this process, each employee may undergo an attribute update with probability $\mu$. Each update alters a single attribute, enabling the exploration of alternative workforce capabilities and organizational arrangements.

The dynamics run for $\tau$ time steps, allowing organizational structures and workforce allocations to stabilize. Once the stationary regime is reached, we average observables over the following $T$ time steps. For each parameter combination ($\theta, \beta$), we perform $\eta$ independent realizations on distinct performance landscapes and report the corresponding ensemble averages. Table~\ref{table: parameters} summarizes all model parameters used throughout this work. Annex A provides the pseudocodes for our dynamics in algorithms \ref{algo:main_loop} and \ref{algo:evolution}.

\begin{figure*}[h]
\centering
\includegraphics[width=1\textwidth]{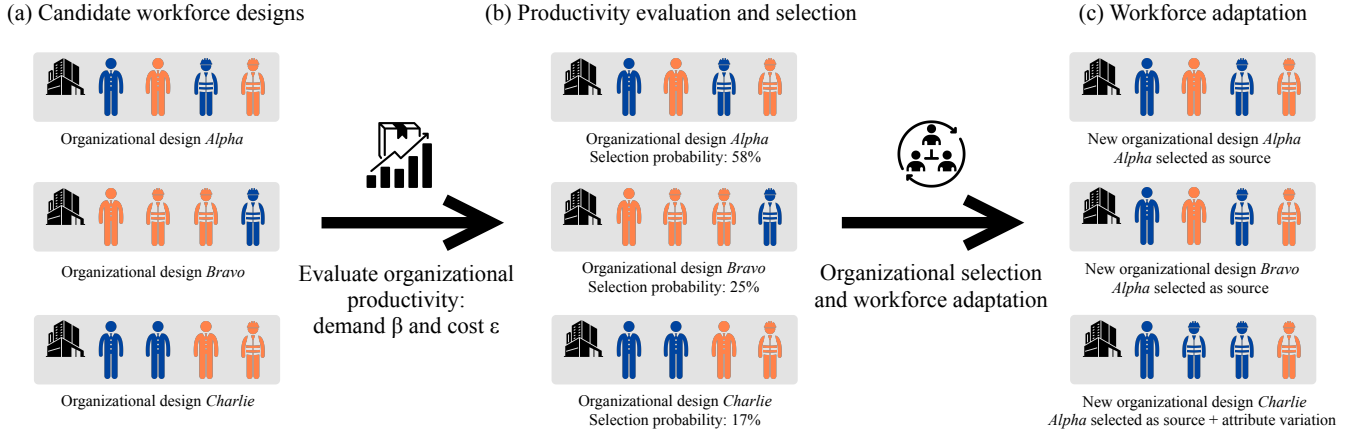}
\caption{Schematic representation of the organizational search and adaptation process for two roles. Employee shape identifies the role performed, with the two shapes representing the two available roles, whereas color indicates workforce strategy: blue represents specialists and orange represents generalists. (a) Candidate workforce designs differ in their employee compositions and attribute profiles. (b) Each design is evaluated according to its organizational productivity, which depends on workforce allocation, which is a function of $\theta$, market demand $\beta$, and the generalist cost $\epsilon$. Designs are selected with probabilities proportional to their productivity, as illustrated by the selection probabilities shown. (c) Selected designs specify the workforce composition of new organizational configurations, while random variations in attributes allow new employee profiles to emerge. The example illustrates one possible update in which design Alpha is selected as the source for the subsequent organizational designs.}
\label{fig:scheme}
\end{figure*}

Figure~\ref{fig:scheme} illustrates the organizational search and adaptation process considered in our model. Starting from a set of candidate workforce designs with two roles, each organization is evaluated according to its productivity, which depends on its workforce composition, which is a function of $\theta$, market demand $\beta$, and the cost $\epsilon$ associated with generalist employees. Colors indicate the employee strategy: blue represents specialists and orange represents generalists. Organizational designs are then selected with probabilities proportional to their productivity, favoring the propagation of more efficient workforce configurations. The selected workforce composition is used to generate new organizational designs, while random variations in attributes allow exploration of alternative employee profiles. Through successive iterations of evaluation, selection, and workforce adaptation, organizations explore the performance landscape and progressively favor workforce designs associated with higher productivity.

\subsection{Topography of the performance landscape}
\label{sec: topography of landscape}

\begin{figure*}[h]
    \centering
    \includegraphics[width=0.82\linewidth]{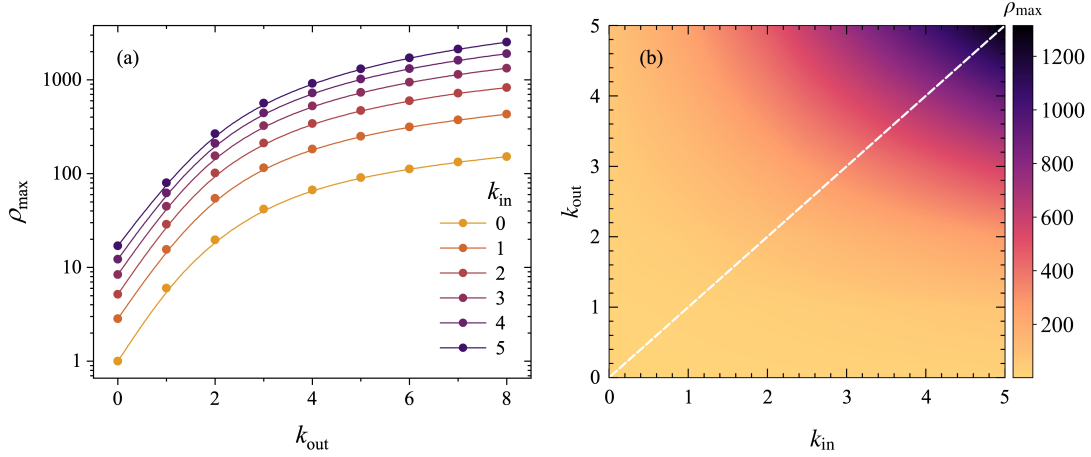}
    \hspace*{+0.5cm}
    %\vspace{-0.5cm}
    \caption{Number of local maxima $\rho_{\mathrm{max}}$ of the productivity landscape as a function of the interdependency parameter $k$ between task attributes. In (a), we present the number of local maxima of the complex landscape versus $k_{\text{out}}$. In (b), we display a heatmap of $\rho_{\mathrm{max}}$ as a funcion of $k_{\text{in}}$ and $k_{\text{out}}$. As $k$ increases, the ruggedness and complexity of the productivity landscape increase, making it more challenging for organizations to find an optimal division of labor. The lines are guides to the eye. In (b), the dashed line is just a reference line for $k_{\text{in}} = k_{\text{out}}$.}
    \label{fig: number of max}
\end{figure*}

Optimization in modern organizations may be understood as an uphill climb in a performance landscape. Its topography is characterized by the number of employee attributes in the roles $1$ and $2$, $N = N_1 + N_2$, and the attribute interdependence parameter, $k = k_{\text{in}} + k_{\text{out}}$. The parameters $k_{\text{in}}$ and $k_{\text{out}}$ capture interdependences within and beyond the role domain, respectively. This allows us to finely tune the effects of connections within a set of skills. 

Within the performance landscape, we define a local optimum for role $r$ as an employee profile $\vec{a}_r^* = \{a_{\ell}^*\}_r$, with $\ell \in N$, whose performance exceeds that of all neighboring profiles obtained through a single attribute modification. In other words, no change to an individual attribute can further improve performance in that task. Unlike the original NK model, where a single fitness landscape characterizes the entire system, the present framework associates a distinct performance landscape with each organizational task, reflecting the possibility that different activities may require different combinations of skills and competencies.

Figure \ref{fig: number of max}(a) displays the number of local optima, $\rho_{\mathrm{max}}$, as a function of $k_{\mathrm{out}}$ for several values of $k_{\mathrm{in}}$. The results show a pronounced increase in $\rho_{\mathrm{max}}$ with both within-task and cross-task interdependencies. For fixed $k_{\mathrm{in}}$, increasing $k_{\mathrm{out}}$ produces a rapid growth in the number of local optima, indicating that stronger coupling between tasks substantially increases the ruggedness of the performance landscape. Likewise, larger values of $k_{\mathrm{in}}$ systematically shift the curves upward, reflecting the additional complexity generated by interactions among attributes associated with the same task. Consequently, highly interdependent organizational structures give rise to a larger number of locally optimal workforce configurations, making the search for globally optimal labor allocations increasingly challenging. In contrast, the limiting case $k_{\mathrm{in}} = k_{\mathrm{out}} = 0$ corresponds to a smooth single-peaked landscape containing only one optimum.

In Fig. \ref{fig: number of max}(b), we show a heatmap representation of the number of local optima $\rho_{\mathrm{max}}$ as a function of both within-task interdependence $k_{\mathrm{in}}$ and cross-task interdependence $k_{\mathrm{out}}$. The color gradient reveals a monotonic increase in $\rho_{\mathrm{max}}$ as either parameter increases, confirming that both forms of interdependence contribute to the ruggedness of the performance landscape despite a small asymmetry. The lowest values occur near the origin, corresponding to weakly coupled tasks and nearly smooth landscapes. In contrast, the upper-right region exhibits the largest number of local optima, indicating that strong interactions both within and across tasks generate highly fragmented performance landscapes with numerous competing solutions. The near-symmetric color distribution about the diagonal $k_{\mathrm{in}} = k_{\mathrm{out}}$ indicates that neither form of interdependence plays a dominant role in generating local optima. Instead, landscape complexity is primarily determined by the total level of interdependence, reaching its highest values when both $k_{\mathrm{in}}$ and $k_{\mathrm{out}}$ are simultaneously large.

\begin{figure*}[!t]
    \centering
    \includegraphics[width=0.88\linewidth]{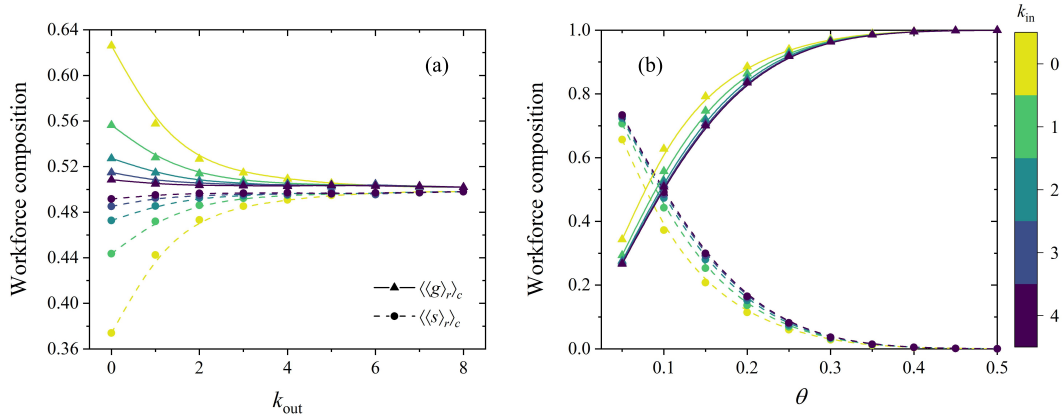}
    \hspace*{-0.5cm}
    \vspace{-0.5cm}
    \caption{Stationary average distribution of specialists and generalist employees for several values of the internal interdependence $k_{\text{in}}$. We show workforce composition versus (a) $k_{\text{out}}$ for the task generalization threshold $\theta = 0.10$, and (b) $\theta$ for $k_{\text{out}} = 0$. For low $k_{\mathrm{in}}$ favors generalists, whereas high $k_{\mathrm{in}}$ favors specialists. As $k_{\mathrm{out}}$ increases, workforce composition converges toward an approximately equal specialist-generalist distribution. As $\theta$ increases, more employees become generalists, eventually leading to a predominantly generalist workforce with reduced functional differentiation. Lines are just guides to the eye.}
    \label{fig: fraction esp/gen}
\end{figure*}

\begin{figure*}[!t]
    \centering
    \includegraphics[width=0.84\linewidth]{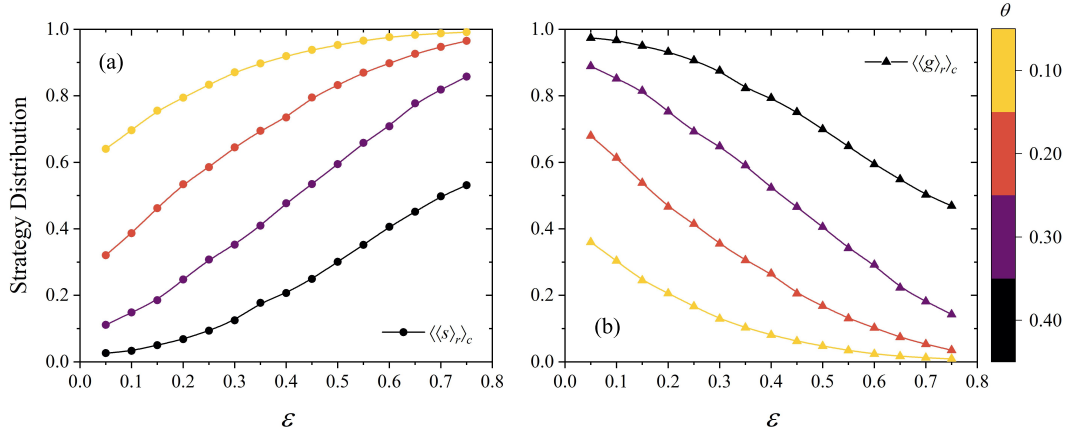}
    %\hspace*{-1cm}
    \vspace{-0.5cm}
    \caption{Stationary average distribution of specialists and generalists as a function of the cost parameter $\epsilon$ for several values of the task generalization threshold $\theta$. Panels (a) and (b) represent the average fractions of specialists and generalists, respectively. Internal and external interdependencies are set to $k_{\text{in}} = 0$ and $k_{\text{out}} = 4$. As $\epsilon$ increases, the selective disadvantage of generalists grows, progressively shifting the workforce toward specialist-dominated configurations. The transition is more abrupt at higher $\theta$, where the initial generalist fraction is larger. Lines are guides to the eye.}
    \label{fig: fraction distribution over c}
\end{figure*}

An important consequence of landscape structure is its impact on workforce allocation and on how interdependencies among employee attributes shape the balance between specialists and generalists in the stationary state. In Fig. \ref{fig: fraction esp/gen}, we investigate the stationary fractions of specialists and generalists across different values of $k_{\mathrm{in}}$, $k_{\mathrm{out}}$, and the generalization threshold $\theta$. Fig. \ref{fig: fraction esp/gen}(a) displays the effect of cross-role interdependencies $k_{\mathrm{out}}$ at fixed $\theta = 0.10$. For weak interdependencies, the workforce composition depends strongly on $k_{\mathrm{in}}$, with low $k_{\mathrm{in}}$ favoring generalists and high $ k_{\mathrm{in}}$ favoring specialists. However, as $k_{\mathrm{out}}$ increases, all curves converge toward an approximately equal distribution of specialists and generalists. This convergence around $50\%$ indicates that strong interactions among roles progressively reduce the distinction between internal and external interdependencies as productivity landscapes become uniform across tasks. As $k_{\mathrm{out}}$ increases, the performance landscape becomes increasingly rugged, giving rise to a larger number of competing attribute combinations capable of producing locally optimal role performance. Consequently, fewer employees exhibit comparable performance across both roles, reducing the fraction of generalists.

Fig. \ref{fig: fraction esp/gen}(b) shows the stationary workforce composition as a function of the generalization threshold $\theta$ for $k_{\mathrm{out}} = 0$. As expected, increasing $\theta$ monotonically increases the fraction of generalists while reducing the fraction of specialists. For small values of $\theta$, only employees with nearly identical performance across roles are classified as generalists, resulting in specialist-dominated organizations. As $\theta$ increases, the performance difference required to define a specialist becomes less restrictive, allowing a growing fraction of employees to qualify as generalists. Eventually, for sufficiently large $\theta$, nearly all employees become generalists, and the workforce loses its functional differentiation. In this scenario, we also note that increasing $k_{\mathrm{in}}$ shifts the transition slightly toward larger values of $\theta$, indicating that stronger within-role interdependencies promote performance differentiation between roles, favoring specialization.

The cost parameter $\epsilon$ associated with generalist employees also critically affects the organizational strategy. Fig. \ref{fig: fraction distribution over c} shows the stationary fraction of specialists and generalists as a function of $\epsilon$ for several values of $\theta$. We observe that increasing the generalist cost drives the system toward specialist-dominated configurations: the fraction of specialists grows monotonically with $\epsilon$, while the complementary fraction of generalists declines accordingly. The sensitivity to $\epsilon$ depends on the generalization threshold. At low values of $\theta$, a large share of employees are already classified as specialists by the landscape itself, and the cost parameter acts primarily on a small generalist minority. As a result, the transition to full specialization is gradual. In contrast, at higher $\theta$ values, the population starts nearly fully generalist, and the increase in $\epsilon$ produces a sharper crossover, reflecting the stronger selective pressure required to overcome the broad generalization window. 

From an organizational perspective, these findings reveal a fundamental trade-off between flexibility and efficiency. Generalists provide adaptability by allowing labor to be dynamically reallocated in response to changing demands, whereas specialists maximize task-specific performance. When the efficiency penalty associated with flexibility becomes sufficiently large, organizations benefit from concentrating employees in specialized roles despite the resulting loss of adaptability. Consequently, $\epsilon$ serves as a key control parameter that governs the balance between organizational resilience and operational efficiency, complementing the effects of landscape structure and workforce heterogeneity discussed previously.

\subsection{Evolutionary dynamics}

\begin{figure*}[h]
    \centering
    \includegraphics[width=0.9\linewidth]{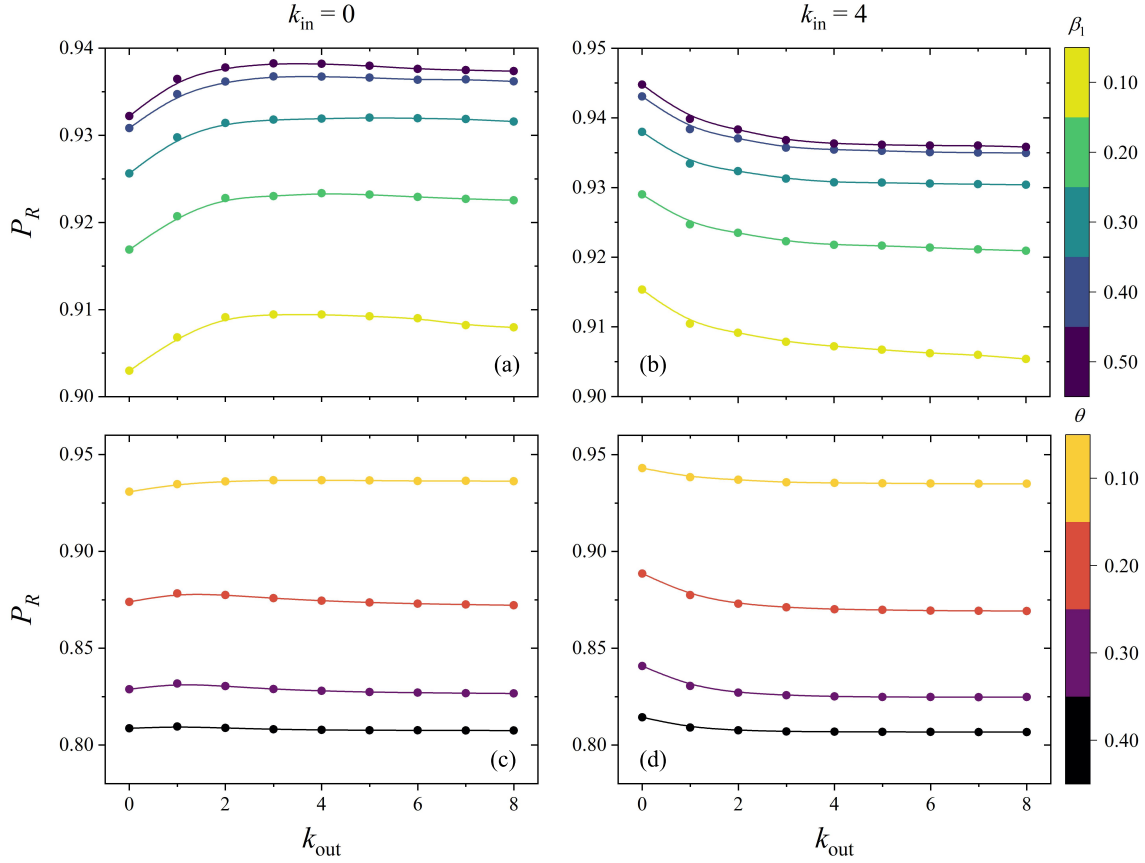}
    \hspace*{-0.2cm}
    \vspace{-0.5cm}
    \caption{Average organization productivity as a function of external task interdependencies for a generalist cost parameter $\epsilon = 0.10$. In panels (a) and (b), we display varying market demands, $\beta_1$, at a fixed generalization threshold $\theta = 0.10$. In panels (c) and (d), we explore a varying generalization threshold, $\theta$, for a fixed market demand $\beta_1 = 0.40$. Internal interdependencies are $k_{\text{in}} = 0$ (left column) and $k_{\text{in}} = 4$ (right column). Moderate cross-task interdependencies enhance productivity when internal interdependencies are weak ($k_{\mathrm{in}}=0$), whereas they reduce productivity when the landscape is already highly rugged ($k_{\mathrm{in}}=4$). More balanced market demands and lower generalization thresholds consistently yield higher organizational productivity. Lines are guides to the eye.}
    \label{fig: productivity comparison}
\end{figure*}

Figure~\ref{fig: productivity comparison} presents the average organizational productivity $P_R$ as a function of the cross-task interdependence $k_{\mathrm{out}}$. The left column corresponds to the absence of within-task interdependence ($k_{\mathrm{in}}=0$), whereas the right column represents a more rugged performance landscape with $k_{\mathrm{in}}=4$. Panels (a) and (b) examine different market demand asymmetries ($\beta_1$) at a generalization threshold $\theta = 0.10$, while panels (c) and (d) investigate different values of the generalization threshold $\theta$ for market demand $\beta_1=0.40$.

For weak within-task interdependence of Figs.~\ref{fig: productivity comparison}(a) and (c), productivity initially increases with $k_{\mathrm{out}}$, reaches a maximum around $k_{\mathrm{out}}\approx2$ to $4$, and subsequently remains nearly constant or decreases slightly. In this regime, moderate cross-task interactions promote workforce differentiation, increasing the fraction of specialists while preserving sufficient flexibility to satisfy market demands. Consequently, organizations benefit from improved task-specific performance without significantly compromising their adaptive capacity.

A qualitatively different behavior emerges when within-task interdependencies are already strong, as in Figs.~\ref{fig: productivity comparison}(b) and (d). In this case, productivity decreases monotonically with $k_{\mathrm{out}}$ for all values of $\beta_1$ and $\theta$. Since the performance landscape is already highly rugged due to the large value of $k_{\mathrm{in}}$, further increases in cross-task coupling primarily increase landscape complexity while reducing the availability of generalists. The resulting loss of workforce flexibility outweighs the productivity gains associated with increased specialization, leading to a gradual decline in organizational performance.

The effects of market demand and the generalization threshold are also clearly visible. More balanced markets ($\beta_1 \approx \beta_2$) consistently achieve higher productivity because labor can be allocated more evenly across roles, whereas stronger demand asymmetries require greater workforce flexibility to maintain efficient allocations. Likewise, increasing the generalization threshold $\theta$ systematically reduces productivity by enlarging the fraction of generalists, whose lower task-specific efficiency introduces an intrinsic productivity penalty. Overall, these results demonstrate that organizational performance is maximized through an appropriate balance between specialization and workforce flexibility, with the optimal balance depending critically on the complexity of the underlying performance landscape.

\begin{figure*}[h]
    \centering
    \includegraphics[width=0.9\linewidth]{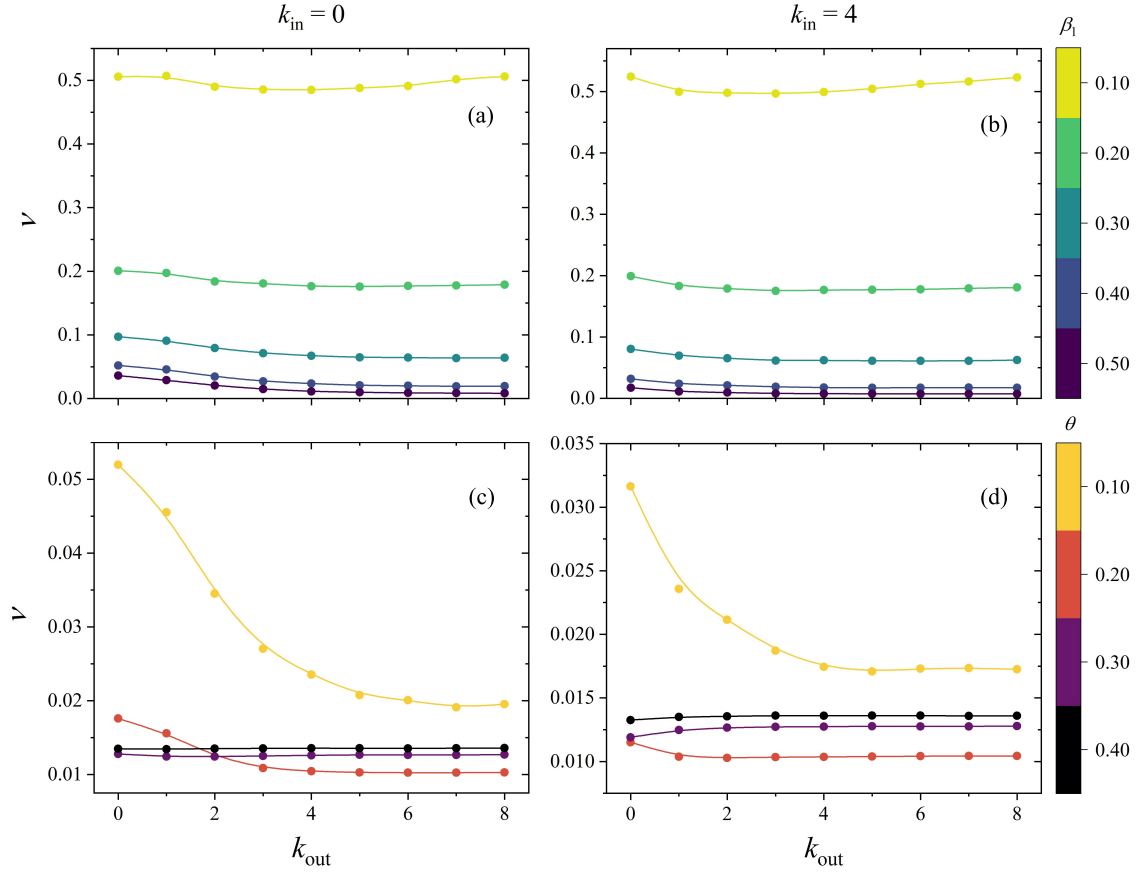}
    \hspace*{0.2cm}
    \vspace{-0.5cm}
    \caption{Deviation from the optimal division of labor as a function of external task interdependencies for a cost parameter $\epsilon = 0.10$. In panels (a) and (b), we show varying market demands, $\beta_1$, for $\theta = 0.10$. In panels (c) and (d), we display the deviation as a function of $\theta$ for $\beta_1 = 0.40$. Internal interdependencies are $k_{\text{in}} = 0$ (left column) and $k_{\text{in}} = 4$ (right column). Cross-task interdependencies improve the alignment between workforce allocation and market demand, especially when internal interdependencies are weak. The smallest deviations occur for balanced markets and intermediate values of the generalization threshold. Lines are guides to the eye.}
    \label{fig: nu comparison}
\end{figure*}

Figure~\ref{fig: nu comparison} presents the deviation from the optimal division of labor, $\nu$, as a function of the cross-task interdependence $k_{\mathrm{out}}$. Recall that $\nu$ measures the largest relative mismatch between the fraction of employees assigned to a task and its corresponding market demand, with smaller values indicating a more efficient workforce allocation. In Fig.~\ref{fig: nu comparison}, the left column corresponds to weak within-task interdependence ($k_{\mathrm{in}} = 0$), whereas the right column represents stronger within-task interactions ($k_{\mathrm{in}} = 4$). Panels (a) and (b) consider different market demand asymmetries ($\beta_1$) at fixed $\theta = 0.10$, while panels (c) and (d) examine different generalization thresholds at fixed demand $\beta_1 = 0.40$.

Figures~\ref{fig: nu comparison}(a) and (b) show that increasing the cross-task interdependence generally reduces the mismatch between workforce allocation and market demand. As $k_{\mathrm{out}}$ increases, the performance landscape becomes more rugged, favoring specialization and reducing the fraction of employees who can perform both roles efficiently. Consequently, organizations naturally allocate employees more consistently based on their task-specific competencies, improving the alignment between workforce composition and market demand. The effect is particularly pronounced in moderately balanced markets, whereas highly asymmetric demands ($\beta_1 = 0.10$) remain difficult to satisfy because the available workforce cannot fully offset the large imbalance in required labor.

Panels (c) and (d) highlight the influence of the generalization threshold for a more symmetric task demand $\beta_1 = 0.40$. For larger values of $\theta$, the deviation from the optimal labor division depends only weakly on $k_{\mathrm{out}}$, since the workforce is predominantly composed of generalists. In contrast, low values of $\theta$ initially produce relatively larger mismatches owing to the balance between specialists and generalists. Increasing $k_{\mathrm{out}}$ progressively reduces deviation $\nu$ by promoting workforce differentiation, allowing employees to become more strongly associated with the role in which they perform best. This improvement is especially evident for $k_{\mathrm{in}} = 0$, and yields similar results for $k_{\mathrm{in}} = 4$ when the workforce is more specialized.

\begin{figure*}[h]
    \centering
    \includegraphics[width=0.88\linewidth]{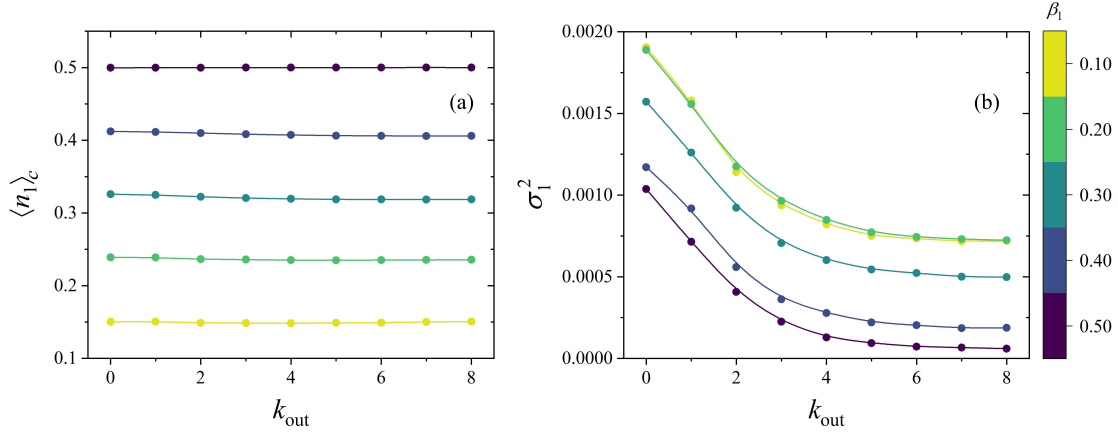}
    \hspace*{-0.2cm}
    \vspace{-0.5cm}
    \caption{Mean and variance of workforce allocation across market demands for a generalization threshold $\theta = 0.10$, cost $\epsilon = 0.10$, and internal interdependence $k_{\text{in}} = 0$. (a) Average fraction of employees performing role $1$, $\langle n_1 \rangle_c$, and (b) the variance of this fraction, $\sigma_1^2$, as a function of external interdependencies, $k_{\text{out}}$, for several values of market demand $\beta_1$. The average workforce allocation remains nearly insensitive to $k_{\mathrm{out}}$, whereas its variance decreases as cross-role interdependencies increase. This indicates that stronger interdependencies primarily stabilize workforce allocation rather than changing its mean value. Lines are guides to the eye.}
    \label{fig: fraction/variance}
\end{figure*}

Figure~\ref{fig: fraction/variance} examines how cross-role interdependencies influence both the average workforce allocation and its fluctuations. Panel (a) presents the stationary average fraction of employees assigned to role $1$, $\langle n_1 \rangle_c$, while panel (b) shows the corresponding variance, $\sigma_1^2$, for different values of the market demand $\beta_1$. The average workforce allocation shown in panel (a) remains remarkably insensitive to the cross-role interdependence $k_{\mathrm{out}}$. For all demand profiles, the stationary fraction of employees assigned to task $1$ closely follows the corresponding market demand, indicating that the labor allocation mechanism successfully adapts the workforce to external requirements independently of the complexity of the underlying performance landscape. Small deviations from the prescribed demand arise only from finite-size fluctuations and the stochastic nature of the organizational dynamics.

Fig. \ref{fig: fraction/variance}(b) shows a different behavior for the workforce fluctuations. The variance $\sigma_1^2$ decreases rapidly as $k_{\mathrm{out}}$ increases for all values of $\beta_1$, eventually approaching a nearly constant value. This reduction indicates that stronger cross-role interdependencies stabilize workforce allocation, making the fraction of employees assigned to each role increasingly reproducible across independent organizational realizations. As the performance landscape becomes more rugged, employees are more strongly differentiated by role-specific competencies, reducing ambiguity in role assignment and suppressing fluctuations in the division of labor. We highlight that the stabilization effect depends on market demand, and more balanced labor markets ($\beta_1 \approx \beta_2$) exhibit lower fluctuations than strongly asymmetric markets.

\section{Final Remarks}

In this work, we explore how division of labor emerges in an organization evolving on a rugged performance landscape. We extend the NK framework to a two-role setting with specialists and generalist agents, and we show that the structure of interdependencies among role attributes strongly shapes the landscape topography and the workforce composition in an organizational design. Specialists perform a single role, whereas generalists can perform multiple roles and are allocated according to organizational demand. Our simulation demonstrates that the organizational composition depends mainly on the interdependence among employee attributes $k$, the generalization threshold $\theta$, and market demand $\beta$. 

An organization's performance depends on finding the right mix of specialists and generalists. Our research suggests that this balance is maintained by two main factors: market demand and the performance landscape, which, in turn, shape employees' responses. We demonstrate that the division of labor is a response to the market, fundamentally determined by how tasks connect and overlap within the organization.

The influence of each employee attribute in a given role depends on its interactions with $k$ other attributes. We divide interactions into two classes: $k_{\rm in}$ with attributes belonging to the same task domain and $k_{\rm out}$ with attributes associated with the other task, satisfying $k = k_{\rm in} + k_{\rm out}$. The generalization threshold $\theta$ determines whether an employee is classified as a specialist or a generalist based on the difference in performance between roles. The market demand for role $r$, $\beta_r$, specifies the target workforce fraction that maximizes organizational productivity. When $k_{\rm in} = 0$, increasing $k_{\rm out}$ promotes the emergence of role-specific attribute combinations, allowing group selection to evolve more skilled specialists while retaining enough generalists to accommodate asymmetric market demands. The resulting increase in task-specific expertise outweighs the efficiency cost associated with workforce flexibility, leading to higher organizational productivity. In contrast, when $k_{\rm in}$ is already high, further increasing $k_{\rm out}$ primarily increases landscape ruggedness rather than improving specialization. 

Therefore, the degree of interdependency can have a beneficial or detrimental role. An intermediate level of interdependency can be beneficial, as it allows group selection to effectively generate attribute configurations that split employees' role performance into two groups: those more efficient at task $1$ and those more efficient at task $2$. When the degree of interdependency is already high, further increases make it difficult for group selection to be effective. As evolutionary dynamics become trapped in numerous local optima and the pool of generalists shrinks, organizations lose their ability to adapt workforce allocation to demand, causing productivity to decline. 

In addition, increasing $\theta$ broadens the range of employees classified as generalists, reducing workforce differentiation and preventing many employees from being assigned exclusively to the role in which they perform best. As a result, a larger fraction of the workforce incurs the efficiency cost of generalization, thereby lowering organizational productivity. At sufficiently large $\theta$, task allocation becomes driven primarily by market demand rather than by employees' comparative advantages, weakening the productivity gains provided by specialization.

The variable $\nu$ measures the maximum gap between the labor division of the current organizational designs and the optimal role allocation. In general, the gap widens as market demand becomes asymmetric, with few specialist employees to meet the increased demand. We observe that the dependence of $\nu$ on $\theta$ is non-monotonic, with the deviation $\nu$ minimized for $\theta = 0.20$, when there is a balance between specialists and generalists to attend the market demand. However, $\theta = 0.40$ increases $\nu$ since each task assignment approaches a random allocation weighted only by the current labor shortage.

Our model, while capturing the essential trade-offs governing the emergence of division of labor, rests on a number of simplifying assumptions whose relaxation may affect the generality of the results. First, the generalist cost parameter $\epsilon$ is uniform and fixed across all employees and tasks, whereas in real organizations, different tasks may impose different efficiency penalties on non-specialists. Second, the attribute exploration probability flips a single randomly chosen attribute, mimicking incremental innovation but excluding mechanisms such as recombination, knowledge transfer, or directed learning, which are important drivers of organizational adaptation. Finally, the restriction to $R = 2$ roles, while sufficient to expose nontrivial phenomena such as the non-monotonic dependence of $\nu$ on $\theta$ and the dual role of interdependencies, limits the complexity of the division-of-labor problem relative to real multi-department organizations. 

Although the present model captures the core mechanisms governing workforce specialization, relaxing some of its simplifying assumptions offers promising directions for future work. Allowing the cost $\epsilon$ to vary across tasks or to depend on the employee's attribute configuration would capture the empirical observation that the efficiency gap between specialists and generalists is itself task-dependent. Another extension is the generalization to $R > 2$ roles, which would introduce richer patterns of inter-task epistasis and allow the study of hierarchical or modular organizational architectures. Furthermore, while the performance landscape for each attribute was generated here using a uniform random distribution, real-world applications typically indicate that the vast majority of attribute combinations are highly unfavorable for a given role, with only a few standout configurations achieving peak efficiency. Consequently, adopting alternative, non-uniform distributions for trait contributions remains an important avenue for creating more realistic performance landscapes.

We also propose that embedding employees in a social or communication network, so that information exchange and coordination costs depend on network topology, could bridge the present framework with the growing literature on organizational network design. Finally, coupling the model to time-varying market demand $\beta_r(t)$ would enable investigation of how organizations dynamically reallocate labor in response to fluctuating or cyclical market conditions, a scenario in which the flexibility afforded by generalists may become substantially more valuable than the present static-demand setting captures.

\section*{Acknowledgements}
The authors acknowledge financial support from Brazilian institutions and funding agencies: UPE, FACEPE (APQ-1129-1.05/24), CAPES, and CNPq (306336/2025-1, 301795/2022-3). PRAC also acknowledges financial support from the National Institute of Science and Technology in Innovative Research in Health Sciences: from Nanotechnology to Artificial Intelligence, sponsored by CNPq (grant no. 408417/2024-2) and FAPESP (grant no. 2025/26818-7). MFBG acknowledges support from the Center and Laboratory for Simulation of Complex Systems, Recife, Brazil, for providing the computational resources and infrastructure necessary for the numerical simulations. We used OpenAI's ChatGPT to assist with language refinement and manuscript editing. The authors solely developed all scientific content, analysis, and interpretation.

\section*{Data availability}
Data will be made available on request.
\clearpage
\appendix
\onecolumn
\subsection*{Appendix A. Evolutionary Dynamics Pseudocodes}

The overall simulation procedure is summarized in Algorithm \ref{algo:main_loop}. For each independent realization, a new NK performance landscape is generated by randomly assigning intra- and inter-role dependencies among employee attributes and constructing the corresponding fitness contributions. The organizational dynamics then evolve in two stages: an initial relaxation period, during which transient effects disappear, followed by a measurement period in which the average organizational productivity and the deviation from the optimal division of labor are sampled. Ensemble averages are finally obtained over all independent realizations.

\begin{algorithm*}[htpb]
\caption{NK Simulation Loop and Landscape Initialization}
\label{algo:main_loop}
\begin{algorithmic}[1]
\State \textbf{Initialize System Parameters:}
\begin{itemize}
  \item $N$ (attributes), $R = 2$ (roles), $k_{\textrm{in}}$ (inner dependencies), $k_{\textrm{out}}$ (outer dependencies)
  \item $C$ (organization designs), $M$ (agents per organization), $\mu$ (exploration rate), $\epsilon$ (generalist cost)
  \item $\tau$ (relaxation time), $T$ (measuring time), $t_{\text{skip}}$ (sampling rate), $\eta$ (samples)
  \item Assign initial macrostate: all individuals set to $\vec{a}_z=\{0\}$
\end{itemize}

\For{$\text{sample} = 1$ to $\eta$}
  \State \textbf{Generate Interdependence Matrix $\mathbf{A}$:}
  \Statex \hskip\algorithmicindent For each attribute $\ell \in \{1, \dots, N\}$, randomly assign $k_{\textrm{in}}$ dependencies from 
  \Statex \hskip\algorithmicindent its own task domain and $k_{\textrm{out}}$ dependencies from the other task domain.
  \Statex \hskip\algorithmicindent Roles $1$ and $2$ are associated to attributes $1,\ldots, N/2$ and $1 + N/2, \ldots, N$, respectively.
  \State \textbf{Generate Random NK Fitness Landscape:}
  \Statex \hskip\algorithmicindent For each attribute $\ell \in \{1, \dots, N\}$ and its dependencies, map all $2^{k_{\textrm{in}} + k_{\textrm{out}} + 1}$ possible binary configurations 
  \Statex \hskip\algorithmicindent to a uniform random fitness contribution $f_\ell \sim \mathcal{U}(0,1)$.
  
  \State Initialize accumulators: $P_{\text{total}} \gets 0$, $\nu_{\text{total}} \gets 0$

  \For{$t = 1$ to $\tau$} \Comment{Relaxation phase}
    \State \Call{TaskAllocation}{$C, M, \mu, \epsilon, \theta, \beta_1, \beta_2$}
    \State \Call{Evolution}{$C, M, \mu, \epsilon, \theta, \beta_1, \beta_2$}
 \Comment{See Algorithm \ref{algo:evolution}}
  \EndFor

  \For{$t = 1$ to $T$} \Comment{Measurement phase}
    \If{$t \bmod t_{\text{skip}} = 0$}
        \State Compute current generation average productivity $P(t)$ and deviation $\nu(t)$
        \State $P_{\text{total}} \gets P_{\text{total}} + P(t)$
        \State $\nu_{\text{total}} \gets \nu_{\text{total}} + \nu(t)$
    \EndIf
    \State \Call{TaskAllocation}{$C, M, \mu, \epsilon, \theta, \beta_1, \beta_2$}
    \State \Call{Evolution}{$C, M, \mu, \epsilon, \theta, \beta_1, \beta_2$}
  \EndFor
\EndFor

\State \textbf{Compute Final Ensemble Averages:}
\State \hskip\algorithmicindent $P_R \gets \frac{P_{\text{total}}}{\eta (T/t_{\text{skip}})}$
\State \hskip\algorithmicindent $\nu \gets \frac{\nu_{\text{total}}}{\eta (T/t_{\text{skip}})}$
\end{algorithmic}
\end{algorithm*}

Algorithm \ref{algo:evolution} details the two main operations performed at every generation. The Task Allocation procedure classifies employees as specialists or generalists based on the performance difference between the two roles, assigns specialists to their best-performing role, and probabilistically allocates generalists to tasks with the greatest labor shortages. The resulting workforce distribution determines the normalized organizational productivity. The Evolution procedure implements a Wright--Fisher group-selection process, in which organizational designs reproduce with probability proportional to their productivity, while attribute exploration introduces small stochastic variations, allowing the evolutionary search to explore new workforce configurations across the performance landscape.

\begin{algorithm*}[htpb]

\caption{Task Allocation and Organization Evolution}
\label{algo:evolution}
\begin{algorithmic}[1]

\Procedure{TaskAllocation}{$C, M, \mu, \epsilon, \theta, \beta_1, \beta_2$}
  \For{$c = 1$ to $C$} \Comment{Evaluate task distribution for each organization}
      \State Initialize counts: $S_{1c} \gets 0$, $S_{2c} \gets 0$, $G_c \gets 0$
      
      \For{$z = 1$ to $M$} \Comment{Classify and allocate specialists}
          \State Compute performance: $F_1 = (2/N) \sum_{\ell=1}^{N/2} f_{\ell}$ and $F_2 = (2/N) \sum_{\ell=1+N/2}^{N} f_{\ell}$
          \If{$|F_1 - F_2| < \theta$}
              \State Agent is a generalist: $G_c \gets G_c + 1$
          \Else
              \State Agent is a specialist
              \If{$F_1 > F_2$}
                  \State Assign to task 1: $S_{1c} \gets S_{1c} + 1$
              \Else
                  \State Assign to task 2: $S_{2c} \gets S_{2c} + 1$
              \EndIf
          \EndIf
      \EndFor
      
      \State Compute specialist fractions: $s_{1c} \gets S_{1c}/M$, $s_{2c} \gets S_{2c}/M$
      
      \State Compute labor shortages: $U_{1c} \gets \max(0, \beta_1 - s_{1c})$, $U_{2c} \gets \max(0, \beta_2 - s_{2c})$
      \State Calculate generalist assignment probabilities: $p_{1c} \gets U_{1c} / (U_{1c} + U_{2c})$ and $p_{2c} \gets 1 - p_{1c}$
      
      \State Initialize generalist assignment counts: $G_{1c} \gets 0$, $G_{2c} \gets 0$
      \For{$z = 1$ to $G_c$} \Comment{Allocate generalists based on shortage}
          \State Draw a uniform random variable $\xi \sim \mathcal{U}(0,1)$
          \If{$\xi \leq p_{1c}$}
              \State Assign generalist to task 1: $G_{1c} \gets G_{1c} + 1$
          \Else
              \State Assign generalist to task 2: $G_{2c} \gets G_{2c} + 1$
          \EndIf
      \EndFor
      
      \State Compute generalist fractions: $g_{1c} \gets G_{1c}/M$, $g_{2c} \gets G_{2c}/M$
      \State Finalize task distribution: $n_{1c} \gets s_{1c} + g_{1c}$, $n_{2c} \gets s_{2c} + g_{2c}$
      
      \State Compute raw productivity: $W_c = n_{1c}^{\beta_1} n_{2c}^{\beta_2} (1 - \epsilon)^{g_{1c} + g_{2c}}$
      \State Compute normalized productivity: $w_c = W_c / (\beta_1^{\beta_1} \beta_2^{\beta_2})$
  \EndFor
\EndProcedure

\vspace{0.3cm}

\Procedure{Evolution}{$C, M, \mu, \epsilon, \theta, \beta_1, \beta_2$}
  \For{$c = 1$ to $C$} \Comment{Wright-Fisher Replication Process}
    \State Select a source group $j$ with probability proportional to its productivity: $P(\text{select } j) = w_j / \sum_{i=1}^C w_i$
    \For{$z = 1$ to $M$} \Comment{Copying process with attribute exploration}
      \State Assign attribute profile: $\vec{a}_{z,c} \gets \vec{a}_{z,j}$
      \State Select one attribute $\ell \in \{1, \dots, N\}$ at random for individual $z$ of company version $c$.
      \State Draw a uniform random variable $\gamma \sim \mathcal{U}(0,1)$
      \If{$\gamma \leq \mu$}
          \State Perform attribute exploration and flip the state of attribute $\ell$: $a_\ell \gets 1 - a_\ell$
      \EndIf
    \EndFor
  \EndFor
\EndProcedure

\end{algorithmic}
\end{algorithm*}

\pagebreak

\twocolumn
\bibliography{References.bib}

\noindent \textbf{Mateus F. B. Granha} is a Ph.D. researcher at the Complexity Science Hub, supervised by Prof. Dr. Peter Klimek (Medical University of Vienna). His current research applies statistical physics and mathematical modeling to investigate the resilience of Austria's food supply chain, with a focus on the beef sector. He holds an M.Sc. in Physics from the Federal University of Pernambuco (UFPE) and a B.Sc. in Materials Physics from the University of Pernambuco (UPE), Brazil. His previous work combined statistical physics, critical phenomena, agent-based modeling, and complex network theory to study how local interactions rooted in social psychology give rise to collective patterns in voting behavior, opinion formation, and financial market fluctuations.\\

\noindent \textbf{Igor V. G. de Oliveira} is a Ph.D. researcher in the physics of nonequilibrium complex systems and stochastic thermodynamics at the University of S\~ao Paulo (USP), supervised by Prof. Dr. Carlos Fiore. He holds an M.Sc. in Physics from the Federal University of Pernambuco (UFPE) and a B.Sc. in Materials Physics from the University of Pernambuco (UPE), Brazil. He also currently works as a scientist at Photrek, Inc., conducting research and developing projects in information thermodynamics and machine learning. He is a member of the Active Inference Institute (AII) and a collaborator at the Center and Laboratory for Simulation of Complex Systems at the University of Pernambuco in Recife, Brazil.\\

\noindent \textbf{Andr\'e L. M. Vilela} is a Full Professor of Physics and the principal investigator of the Center and Laboratory for Simulation on Complex Systems at the University of Pernambuco (UPE), Brazil. He serves as Associate Editor of the Humanities and Social Sciences Communications. He has investigated the Physics of Complex Systems, with a particular interest in the dynamics of interacting agent-based models, statistical mechanics, phase transitions, critical phenomena, and finite-size scaling analysis associated with opinion dynamics, financial markets, and complex network theory, including applications to data characterization, network resilience, and decentralized decision-making technology and artificial intelligence. His research focuses on uncovering the underlying mathematical mechanisms that drive the behavior of interacting agents, or constituents, within the complex network framework and on how their behavior promotes active collective phenomena in social, technological, and economic systems. \\

\noindent \textbf{Chao Wang} is a Full Professor and Doctoral Supervisor at the School of Economics and Management, Beijing University of Technology (BJUT), China. He received his Ph.D. in Management Science and Engineering from Beijing Jiaotong University and was a Postdoctoral Research Fellow in the Department of Physics at Boston University, working with the late Prof. H. Eugene Stanley. He was also a visiting scholar at Purdue University and Tsinghua University. His research integrates complex network theory with resource and environmental economics, focusing on the structural evolution of global trade networks, critical mineral and power battery supply chains, and supply chain resilience and security. His work has appeared in PNAS, Omega, Applied Energy, Renewable and Sustainable Energy Reviews, Resources, Conservation and Recycling, and International Journal of Production Economics, among others. He serves as Co-Editor-in-Chief of Sustainable Futures (Elsevier) and Associate Editor of the International Journal of Logistics Research and Applications.\\

\noindent \textbf{Paulo R. A. Campos} is a Full Professor of Physics at the Federal University of Pernambuco (UFPE), Recife, Brazil. His research lies at the interface of statistical physics, evolutionary biology, ecology, and complex systems, with particular emphasis on stochastic models of population dynamics and evolution. He has investigated evolutionary adaptation and rescue, fitness landscapes, population genetics, biodiversity, ecological interactions, spatially structured populations, and the effects of environmental variability on evolutionary and ecological processes. His work also encompasses agent-based and network models of collective behavior, cooperation, cultural dynamics, and social organization. His research focuses on identifying the statistical and mathematical mechanisms underlying adaptation, persistence, diversity, and emergent collective behavior in biological and complex systems.\\

\end{document}